\documentclass[journal,comsoc]{IEEEtran}

\usepackage[T1]{fontenc}

\usepackage{cite}
\ifCLASSINFOpdf
  \usepackage[pdftex]{graphicx}
  \graphicspath{{../pdf/}{../jpeg/}{./graphics/}}
  \DeclareGraphicsExtensions{.pdf,.jpeg,.png}
\else
  \usepackage[dvips]{graphicx}
  \graphicspath{{../eps/}}
  \DeclareGraphicsExtensions{.eps}
\fi
\usepackage{amsmath}
\usepackage[cmintegrals]{newtxmath}
\usepackage{bm}
\ifCLASSOPTIONcompsoc
  \usepackage[caption=false,font=normalsize,labelfont=sf,textfont=sf]{subfig}
\else
  \usepackage[caption=false,font=footnotesize]{subfig}
\fi
\usepackage{url}
\usepackage{iitem}
\usepackage{tikz}
\usetikzlibrary{positioning}
\usetikzlibrary{patterns}
\usetikzlibrary{backgrounds}
\usepackage{siunitx}
\newcommand{\refval}{s_\textsc{rx}}
\newcommand{\dimnum}{n_\mathrm{d}}
\DeclareMathOperator{\round}{round}

\newcommand{\interferencethreshold}{\xi_\mathrm{ISI}}
\newcommand{\detectionthreshold}{\xi_\mathrm{det}}

\newcommand{\tmax}{t_\mathrm{max}}

\newcommand{\pmax}{p_\mathrm{max}}
\newcommand{\minimumresource}{m_\mathrm{min}}
\newcommand{\maximumresource}{m_\mathrm{max}}
\newcommand{\copyablelabel}[1]{\label{#1}}

\newcommand{\maxsignal}{peak detection value}
\newcommand{\detectionvalue}{detection value}

\newcommand{\figscal}{0.8}
\newcommand{\changed}[2][]{{#2}}

\begin{document}
\bstctlcite{IEEEexample:BSTcontrol}
\title{Pulse Shaping for MC via Particle Size}
\author{%
    Wayan Wicke,~\IEEEmembership{Graduate Student Member,~IEEE,}
    Rebecca C.\ Felsheim,
    Lukas Brand,~\IEEEmembership{Graduate Student Member,~IEEE,}
    Vahid Jamali,~\IEEEmembership{Member,~IEEE,}
    Helene M.\ Loos,
    Andrea Buettner,
    Robert Schober,~\IEEEmembership{Fellow,~IEEE,}%
    \thanks{%
        This work was supported in part by the Emerging Fields Initiative (EFI) of the Friedrich-Alexander-Universit\"at Erlangen-N\"urnberg (FAU) and the STAEDTLER-Stiftung.
        \textit{(Corresponding author: Wayan Wicke.)}
    }%
    \thanks{%
        W.~Wicke, R.~C.~Felsheim, L.~Brand, and R.~Schober are with the Institute for Digital Communications, Friedrich-Alexander-Universit\"at Erlangen-N\"urnberg (FAU), 91058 Erlangen, Germany (e-mail: wayan.wicke@fau.de; rebecca.felsheim@fau.de; lukas.brand@fau.de; robert.schober@fau.de).
    }%
    \thanks{%
    V.~Jamali is with the Department of Electrical Engineering and Information Technology, 64283 Darmstadt, Germany (e-mail: vahid.jamali@tu-darmstadt.de).
    }%
    \thanks{%
        H.~M.~Loos and A.~Buettner are with the Chair of Aroma and Smell Research, Friedrich-Alexander-Universit\"at Erlangen-N\"urnberg (FAU), 91054 Erlangen, Germany, and the Fraunhofer Institute for Process Engineering and Packaging IVV, 85354 Freising, Germany.%
    }
    \vspace*{-10mm}
}


\maketitle
\begin{abstract}
In molecular communication (MC), combining different types of particles at the transmitter is a degree of freedom which can be utilized to improve performance.
In this paper, we address the problem of pulse shaping to simplify time synchronization requirements by exploiting and combining the received signal characteristics of particles of different sizes.
\changed{In particular, we optimize the mixture of particles of different sizes used for transmission in order to support a prescribed detection time period for on-off keying, guaranteeing on average 1) a sufficiently large received signal if a binary one is transmitted, and 2) a low enough received signal if a binary zero is transmitted even in the presence of inter-symbol interference.}
For illustration, we consider an optimization problem based on a free space diffusion channel model.
It is shown that there is a tradeoff between the maximum feasible detection duration and the \maxsignal{} for different particle sizes from the smallest particle size enabling the largest detection duration to the largest particle size minimizing the \maxsignal{} at the expense of a limited detection duration.
\end{abstract}

\IEEEpeerreviewmaketitle

\section{Introduction}
Molecular communication (MC) is the prime means of information transmission in natural systems, be it within organisms or between individuals or species.
The diversity of information representation, the diverse transmission media, and the range of distances covered make MC also a promising option for technical systems~\cite{jamali_channel_2019}.

Research on MC usually considers a single type of particle~\cite{jamali_channel_2019} or multiple types of particles which can be detected independently~\cite{kuran_survey_2021}.
\changed{%
Using multiple particle types is a common strategy in natural systems, e.g., the chemical signals used by insects~\cite{wyatt_pheromones_2014} are typically comprised of chemical mixtures, which are detected jointly by the receiver.%
} 
In natural MC systems, the constituents may also react or degrade and bring forth derivatives or artifacts.
Those, again, can provide information on, e.g., distance, environmental factors, or the impact of the gaseous or liquid media.

\changed{For artificial MC systems, even if just a single type of particle is used, guidelines are needed for choosing the type of information particle.}
In fact, particles can be designed to have different sizes, shapes, surface coatings, reaction constants, chemical stability, etc.~\cite{zaloga_development_2014}, which affect the communication in different ways.
These properties can be used to design received pulse shapes with different pulse heights, widths, and rates of decay which affect the signal strength, robustness to time synchronization offsets, and inter-symbol interference (ISI), respectively.
While time synchronization can be established in principle, accurate synchronization entails a large signaling overhead~\cite{jamali_symbol_2017}.
Thus, in practice, potentially detrimental timing errors remain.
Hence, it is desirable that the received pulse shape has a relatively wide peak within one symbol interval for robustness against timing errors and a fast decay to avoid ISI.
Since these requirements are contradictory, design guidelines are needed for choosing suitable particles as information carriers.

\changed{%
In MC systems, the end-to-end channel impulse response (CIR) includes the impact of the release, propagation, and reception mechanisms.
For pulse shape design in MC, i.e., the deliberate design of the parameters affecting the CIR, the times of release have been optimized~\cite{cao_diffusive_2019}, reactive particle types have been used for transmission (acid and base chemicals)~\cite{farsad_molecular_2016}, and the signals received for two different molecule types have been combined \cite{tepekule_novel_2015}.
Moreover, pulse shaping by optimizing one of two particle sizes in a two-transmitter-one-receiver scenario has been considered in \cite{angjo_asymmetrical_2022}.
However, using multiple differently sized particles of the same type and combining their received signal characteristics for pulse shaping has not been studied in the literature, yet.}

The contributions of this paper can be summarized as follows.
\begin{enumerate}
	\item We propose the idea of pulse shaping \changed{of the end-to-end received signal} by combining particles of different sizes and observing a linear superposition of the individual responses of the MC channel to these particles.
	\item For free space diffusion, we optimize the mixture for minimization of the \maxsignal{} while abiding to a detection lower bound and an ISI upper bound for detection of binary symbols $1$ and $0$ for on-off keying (OOK), respectively.
	Minimizing the \maxsignal{} saves resources at the transmitter and reduces potential interference with other communication links.
	\item We provide insight regarding feasible detection durations and the selection of particle sizes.
\end{enumerate}


\section{System Model}
\label{sec:system model}
%
We consider an unbounded environment comprising a point transmitter and a spherical transparent receiver of volume $V$.
The distance between transmitter and receiver is denoted as $d$.
We assume that the transmitter can release multiple spherical particles instantaneously.
The particles released by the transmitter can have $M$ different sizes with radius $R_i$ and are subject to diffusion characterized by diffusion coefficient $D_i$, $i\in\{1,2,\dots,M\}$.
The number of particles having the same diffusion coefficient $D_i$ released by the transmitter is denoted as $n_i$.
For convenience, we consider a reference particle size with radius $R_0$ and with reference diffusion coefficient $D_0$.
Then, for particles of radius $R_i$, the relative particle size is defined as $\rho_i=R_i/R_0$, i.e., $\rho_0=1$ for the reference particle size.
By the Einstein relation, the diffusion coefficient $D_i$ for relative particle size $\rho_i$ can be written as $D_i=D_0/\rho_i$~\cite{jamali_channel_2019}.
\changed{A dilute mixture of particles is considered where no interaction among particles either of the same or of different sizes is assumed, i.e., the particles move unobstructed and independent of each other.
For example, in medical applications, magnetic nanoparticles can be produced in different sizes and synthesized with a suitable coating such as lauric acid to avoid agglomeration~\cite{zaloga_development_2014}.}

In general, differently sized particles may generate different \changed{\emph{\detectionvalue{}s}} depending on the physical particle detection mechanism.
For example, when detecting the iron content of magnetic nanoparticles~\cite{zaloga_development_2014}, the \detectionvalue{} may depend on their volume or surface area depending on how the iron molecules within one particle are distributed, e.g., uniformly within the particle volume or on its surface.
On the other hand, the iron content per particle may also be independent of the particle size, e.g., when the particles have a fixed iron core and a coating of non-magnetic material of variable size.
With this motivation, we consider a general \detectionvalue{} metric for detecting particles of different sizes.
In particular, denoting the \detectionvalue{} for one single particle of the reference size as $\refval$, we assume the \detectionvalue{} for one single particle of size $R_i$ is given by $\refval\rho_i^{\dimnum}$, where $\dimnum$ is the number of spatial dimensions in which the \detectionvalue{} of a particle scales with its size.
For $\dimnum=0$, the \detectionvalue{} for a particle does not scale with its size, for $\dimnum=2$, it scales with its surface area, and, for $\dimnum=3$, it scales with its volume.
The overall received signal is then obtained as the sum of the \detectionvalue{}s of the individual particles.

For modulation, the transmitter uses OOK with transmit symbols $a_k\in\{0,1\}$ and symbol duration $T$.
For transmitting $a_k=1$ in symbol interval $k$, the transmitter releases a given mixture containing particles of different sizes at the beginning of each time interval at time $kT$.
For transmitting $a_k=0$, the transmitter does not release particles and remains silent.
For detection, the receiver is assumed to process finitely many samples in each symbol interval, e.g., a single sample per symbol interval. 
\changed{However, no particular detection algorithm is assumed in this work to keep the pulse shape design general and not specific to a certain type of detector}.

\section{Problem Formulation}
\label{sec:analysis}
In this section, we introduce the proposed optimization framework and analyze the special case of using just a single particle size after introducing our notation.
For a given symbol interval $k$, samples of the average received signal (normalized by $\refval$), $r(t)$, are gathered in vector $\bm{r}$ with samples $r_l=r(kT+l\Delta t)$ for $l=0,1,\dots,L-1$ taken in time intervals $\Delta t=T/L$, where $L$ is the number of samples per symbol.
To make detection robust to time offsets \changed{and ISI}, we assume that for detection of symbol $a_k=1$, the \changed{average} received signal should satisfy $r_l\geq\detectionthreshold$ and, for detection of $a_k=0$, it should satisfy $r_l<\interferencethreshold$, where $l_0\leq l< l_0+L_0$.
\changed{Here, $\detectionthreshold$ and $\interferencethreshold$ are \emph{design parameters} which will be referred to as detection threshold and ISI threshold, respectively.}
\changed{Moreover, the arbitrary sampling instance offset $l_0$ is also a design parameter, while $L_0$ specifies the number of samples for which detection is reliably feasible and can be chosen depending on the expected detection time offsets.}
Assuming an arbitrary given symbol-by-symbol detection \changed{scheme, e.g., based on the detection threshold}, the detection threshold controls the minimum \changed{average} required signal strength for detecting binary $1$ and the ISI threshold controls the maximum \changed{average} permissible signal strength due to ISI for detecting binary $0$.
\changed{Thereby, thresholds $\detectionthreshold$ and $\interferencethreshold$ control the symbol error rate while parameter $L_0$ makes detection robust to time synchronization errors.}
The signal values in the remaining $L-L_0$ samples per symbol interval are not constrained\changed{, i.e., for these samples reliable detection might not be feasible}.
For convenience, we also define the corresponding detection starting time $t_0=l_0\Delta t$ and detection time duration $T_0=L_0\Delta t$.

\subsection{Mathematical Model}
The received signal is stochastic in nature due to the independent reception of each particle.
Nevertheless, the detection performance generally depends on the average received signal which we will consider in the following.
The average received signal (averaged over the receiver counting noise) for the assumed OOK modulation can be written as
\begin{equation}
\copyablelabel{eq:PAM signal}
	r(t) = \sum_{k=-\infty}^\infty a_k h(t-kT)
\end{equation}
where $h(t)$ is the (average) received pulse shape for the given particle mixture at the transmitter.
The channel CIR for particles with diffusion coefficient $D$ is denoted as $p(t; D)$.
For the assumed transparent receiver, it describes the probability to observe one particle of diffusion coefficient $D$ released by the transmitter at time $t=0$ within the receiver volume at time $t$.
The average number of received particles of size $R_i$ can then be written as $n_i p(t; D_i)$.
Hence, with the \detectionvalue{} for detecting a particle of size $\rho_i$, $\refval\rho_i^{\dimnum}$, the average \detectionvalue{} for $n_i$ released particles of size $\rho_i$ can be expressed as $\refval\rho_i^{\dimnum}\cdot n_i p(t; D_i)$.
Consequently, the overall average pulse shape (normalized by $\refval$) in \eqref{eq:PAM signal} is the superposition of the average \detectionvalue{}s of all particle types and can be expressed as
\begin{equation}
\copyablelabel{eq:pulse shape}
	h(t) = \sum_{i=1}^M \rho_i^{\dimnum} n_i p(t;D_i),
\end{equation}
which is equal to $r(t)$ for $a_0=1$ and $a_k=0$ for $k<0$ and $k>0$.
Similarly, we define the ISI pulse shape $h_\mathrm{r}(t)$ which is obtained from $r(t)$ in \eqref{eq:PAM signal} by setting $a_0=0$ and $a_k=1$ for $k<0$.

In the following, we assume free space diffusion \changed{and a transparent receiver}.
In this case, the CIR for a particle with diffusion coefficient $D$ can be written as~\cite{jamali_channel_2019}
\begin{equation}
\copyablelabel{eq:CIR}
	p(t; D) = \frac{V}{\left(\sqrt{4\pi D t}\right)^3}\cdot \exp\bigg(-\frac{d^2}{4 D t}\bigg), t\geq 0,
\end{equation}
and $p(t; D)=0, t<0$.
\changed{%
Moreover, we define the worst-case ISI signal resulting from infinitely many consecutive releases as follows%
}%
\begin{equation}
\copyablelabel{eq:ISI CIR}
	p_\mathrm{r}(t; D) = \sum_{k=1}^\infty p(t+kT; D)\changed{, t\geq 0}.
\end{equation}
%
We further assume the samples of the CIRs are known and gathered in column vector $\bm{p}_i=(p_{i,0},p_{i,1},\dots, p_{i,L-1})$, where $p_{i,l} =p(l\Delta t; D_i)$.
In the same way, we also define a vector for the \changed{ISI signal} as $\bm{p}_{\mathrm{r},i}=(p_{\mathrm{r},i,0}, p_{\mathrm{r},i,1},\dots,p_{\mathrm{r},i,L-1})$ with $p_{\mathrm{r},i,l}=p_\mathrm{r}(l\Delta t; D_i)$.
For convenience, we define CIR matrix $\bm{P}=[\bm{p}_0, \bm{p}_1, \dots, \bm{p}_{M-1}]$ with CIR time samples $\bm{p}_i$ for particles with diffusion coefficient $D_i$ and $\bm{P}_\mathrm{r}=[\bm{p}_{\mathrm{r},0},\bm{p}_{\mathrm{r},1},\dots,\bm{p}_{\mathrm{r},M-1}]$ as the corresponding \changed{ISI signal} matrix.

For convenience, we define $m_i=\rho_i^{\dimnum} n_i$ and $\bm{m} = (m_1, m_2,\dots,m_M)$ representing the \maxsignal{} when detecting all particles of relative size $\rho_i$.
With this notation, the worst-case received signal vector for detecting $a_0=1$ (i.e., the smallest received signal obtained for $a_k=0, \forall k<0$) can then be written as $\bm{r}=\bm{h}$ where vector
\begin{equation}
\copyablelabel{eq:vector received signal}
	\bm{h}=\bm{P}\bm{m},
\end{equation}
i.e., vector $\bm{h}$ contains the samples of pulse shape $h(t)$.
Similarly, the worst-case received signal vector for detecting $a_0=0$ (i.e., the largest received signal obtained for $a_k=1,\forall k<0$) can be written as $\bm{r}=\bm{h}_\mathrm{r}$ where
\begin{equation}
\copyablelabel{eq:vector received signal ISI}
	\bm{h}_\mathrm{r}=\bm{P}_\mathrm{r}\bm{m}
\end{equation}
contains the samples of ISI pulse shape $h_\mathrm{r}(t)$.

Moreover, the resulting \maxsignal{} (when all particles were detected at the same time) can be written as
\begin{equation}
\copyablelabel{eq:particle budget}	
N = \bm{1}^\intercal\bm{m},
\end{equation}
where $\bm{1}^\intercal$ is a row vector containing only ones.

\subsection{Optimization Problem}
\begin{figure}[t]
	\centering
	\includegraphics[scale=\figscal]{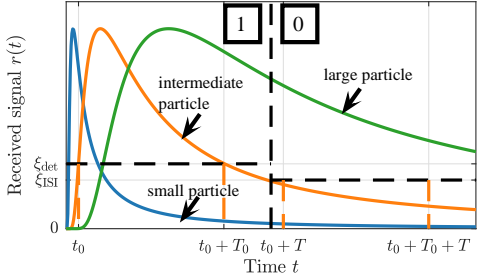}
	\vspace*{-3mm}
    \caption{%
    Problem illustration.
    Received signal for $a_0=1$ and $a_1=0$ ($a_k=0$ for $k<0$) for small, intermediate, and large particle sizes, assuming the same \maxsignal{}.
    Dashed black lines mark the detection and ISI thresholds and the symbol duration, respectively.
    Dashed orange lines mark the largest possible detection periods for the intermediate particle size.
    \vspace*{-5mm}
    }
    \copyablelabel{fig:system model}
\end{figure}
The basic idea behind the proposed design is illustrated in Fig.~\ref{fig:system model}.
There, the received signal is plotted for the transmitted sequence $a_0=1$ and $a_1=0$ with no transmission before or afterwards, i.e., no ISI for symbol interval $k=0$ but relatively large ISI for symbol interval $k=1$.
Hence, for detection, we require the received signal to be larger than $\detectionthreshold$ for $t_0\leq t\leq t_0+T_0$ and smaller than $\interferencethreshold$ for $T+t_0\leq t\leq T+t_0+T_0$, where $T_0$ is the detection duration.
Larger values of $T_0$ indicate more robustness towards detection time offsets and $t_0$ can be adjusted accordingly.
We show the average received signal for three particle sizes, labeled small, intermediate, and large but for the same \maxsignal{}.
For the small particle size, the received signal lies above the given detection threshold only for a small duration of time but the ISI caused is also small and in fact below the ISI threshold for any $t_0\geq T$.
For the intermediate particle size, the CIR lies above the detection threshold for a longer time duration but at the expense of more but still admissible ISI.
The large particle size would offer an even longer detection duration but is not feasible since the ISI is above the ISI threshold for any $t_0\geq T$.
In summary, there is a tradeoff between the possible detection window size (indicating robustness to detection time offsets) and the ISI in terms of particle size.
For a given \maxsignal{}, there exists an intermediate particle size with maximal detection window size abiding to the detection and ISI constraints.
By adapting the detection and ISI thresholds, this tradeoff can be adjusted to favor larger or smaller particles.
Finally, by combining particles of different sizes, there are additional degrees of freedom which can be exploited for receive pulse shaping.

In the following, for a resource efficient design minimizing potential interference with other communication links, we consider the minimization of the \maxsignal{} \changed{in \eqref{eq:particle budget}} via the optimization of the particle mixture subject to detection constraints for the worst-case received signals in \eqref{eq:vector received signal} and \eqref{eq:vector received signal ISI}.
Thus, we can pose the following optimization problem
\begin{IEEEeqnarray}{lrCl}
	\underset{m_i=\rho_i^{\dimnum}n_i,\,l_0}{\mathrm{minimize}}\; &\bm{1}^\intercal\bm{m} \IEEEyesnumber\copyablelabel{eq:optimization}\IEEEyessubnumber*\\
	\text{subject to}\; &\bm{P}\bm{m}&\succeq &\bm{w}_\mathrm{det}\\
	&\bm{P}_\mathrm{r}\bm{m} &\prec& \bm{w}_\mathrm{ISI}\\
	&\bm{m}&\succeq &\bm{0}
\end{IEEEeqnarray}
where $n_i$ for $i=1,2,\dots,M$ need to be integers and $\succeq$ and $\prec$ denote elementwise inequalities.
For the detection constraint window, we have $\bm{w}_\mathrm{det}=(w_{\mathrm{det},0}, w_{\mathrm{det},1},\dots,w_{\mathrm{det},L-1})$ where $w_{\mathrm{det},l}=\detectionthreshold$ if $l_0\leq l< l_0+L_0$ and $0$ otherwise.
Moreover, for the ISI constraint window, we have $\bm{w}_\mathrm{ISI}=(w_{\mathrm{ISI},0}, w_{\mathrm{ISI},1},\dots,w_{\mathrm{ISI},L-1})$ where $w_{\mathrm{ISI},l}=\interferencethreshold$ if $l_0\leq l< l_0+L_0$ and $\infty$ otherwise.
We note that both $\bm{w}_\mathrm{det}$ and $\bm{w}_\mathrm{ISI}$ depend on $l_0$.

In general, the optimization problem in \eqref{eq:optimization} is difficult to solve because 1) $m_i$ can only assume values from a discrete set of real numbers ($n_i$ is an integer) and 2) offset $l_0$ is an integer value as well influencing the constraint window position.
However, assuming the elements of $\bm{m}$ could assume arbitrary real values, the problem in \eqref{eq:optimization} is a linear program with respect to $\bm{m}$ which can be efficiently solved numerically using, e.g., CVX, for a given time offset $l_0\in[0,L-L_0]$.
To solve the joint problem for $\bm{m}$ and $l_0$, we loop over scalar parameter $l_0$ and finally choose the solution with the smallest \maxsignal{}, i.e., we perform a one-dimensional search over scalar variable $l_0$ which entails a low computational complexity.
Then, from the found solution for $\bm{m}$, we can obtain an approximate solution to \eqref{eq:optimization} by scaling and rounding as $m^*_{i}=\rho_i^{\dimnum} \round\left(m_i/\rho_i^{\dimnum}\right)$.
We note that in general, because of the rounding, this is a suboptimal solution.
Nevertheless, in the following, we assume a reasonable design is obtained in this way since $m^*_i$ is close to $m_i$ found by optimization if $m_i/\rho_i^{\dimnum}$ is large.

\subsection{Single Particle Size}
As a benchmark for our optimization, we consider the case where just a single particle size can be selected.
In this case, we consider the two constraints $m p(t;D)\geq\detectionthreshold$ for $t_0\leq t\leq T$ and $m p_\mathrm{r}(t;D)\leq\interferencethreshold$ for $t\geq t_0$.
Even for this benchmark, a solution to \eqref{eq:optimization} is not straightforward since the constraints have a joint dependency on $t_0$ (i.e., $l_0$) and $m$.
In addition, it is not straightforward to determine which particle size should be preferred when multiple sizes are possible.
Nevertheless, bounds can be derived as is done in the following.

First, let us analyze the CIR in \eqref{eq:CIR} for one single particle size.
This CIR is unimodal with its peak position at time $\tmax=d^2/(6D)$~\cite{jamali_channel_2019}.
Consequently, the peak value of the CIR is obtained by substituting the peak time $\tmax$ in \eqref{eq:CIR} as
\begin{equation}
	\pmax = \frac{V}{d^3\left(\sqrt{\frac{2}{3}\pi\mathrm{e}}\right)^3}
\end{equation}
which is independent of $D$.
Thereby, a lower bound on the minimum required \maxsignal{} for any particle size can be specified as
\begin{equation}
\copyablelabel{eq:minimumresource}
	\minimumresource = \frac{\detectionthreshold}{\pmax}
\end{equation}
which is obtained by scaling the peak of the pulse shape to the detection threshold.
However, by our constraints, $\minimumresource$ is only applicable if we have $\minimumresource p_\mathrm{r}(\tmax;D)<\interferencethreshold$.
Otherwise, there is no solution.
If we assume that $m>\minimumresource$ is feasible, there will be two time instances $t_1<t_2$, where $t=t_1\leq\tmax$ and $t=t_2\geq \tmax$, for which $m p(t;D)=\detectionthreshold$.
Then, for such a given relative cost $m$, the feasible detection duration is $T_0=t_2-t_1$, where we can assume that $t_2<T$ since otherwise $p_\mathrm{r}(t_1)\geq \interferencethreshold$.

In the following, we consider the limiting case of very small particles, motivated by the following observation.
When selecting among different particle sizes, from Fig.~\ref{fig:system model}, it can be concluded that smaller particles support a smaller detection window compared to larger particles.
This is true when considering the same \maxsignal{} among the considered particles.
However, as can also be seen from Fig.~\ref{fig:system model}, the ISI for smaller particles is reduced compared to larger particles.
Hence, at the expense of a larger \maxsignal{}, a longer detection duration is possible for smaller particles compared to larger particles.
In this way, the small particle limit can be considered to enable the longest detection duration.
Mathematically, the small particle limit can be expressed by the limit $\rho\to0$ which corresponds to $D\to\infty$.
In this case, the peak position becomes $\tmax=0$ and hence $t_0=t_1=0$ can be chosen and thus $T_0=t_2$.
Furthermore, the exponential terms in \eqref{eq:CIR} can be neglected and equated to $1$.
Thus, there is only one solution for $m p(t;D)=\detectionthreshold$ for a given relative cost $m$, which can be obtained as
\begin{equation}
\copyablelabel{eq:smallparticledetectionduration}
	T_0 = \frac{1}{4\pi D} \left(\frac{m V}{\detectionthreshold}\right)^{\frac{2}{3}}.
\end{equation}
By using the ISI constraint, the maximum relative cost can be written as
\begin{equation}
\copyablelabel{eq:smallparticlescaling}
	\maximumresource = \frac{\interferencethreshold}{p_\mathrm{r}(0;D)}
\end{equation}
which is obtained by scaling the ISI signal at time $t_0=t_1=0$ to the ISI threshold.
Finally, we can derive an upper bound for the detection duration as
\begin{equation}
\copyablelabel{eq:smallparticlebound}
	\frac{T_{0,\mathrm{max}}}{T} = \left(\frac{\interferencethreshold}{\detectionthreshold \zeta(3/2)}\right)^\frac{2}{3}
\end{equation}
by substituting \eqref{eq:smallparticlescaling} in \eqref{eq:smallparticledetectionduration} and using $p_\mathrm{r}(0; D)=V\cdot \zeta(3/2)/\sqrt{4\pi DT}^3$ which can be obtained by simplifying the exponential term in \eqref{eq:ISI CIR} to $1$ and using $\sum_{k=1}^\infty k^{-3/2}=\zeta(3/2)$ where $\zeta(\cdot)$ is the Riemann zeta function.

\section{Numerical Results}
\label{sec:results}
\changed{For illustration, we consider the following exemplary parameters adapted from \cite{zaloga_development_2014}, unless specified otherwise:
Reference particle radius $R_0=\SI{25}{\nano\meter}$ with reference diffusion coefficient $D_0=\SI{8e-12}{\meter^2\per\second}$ and $n_\mathrm{d}=3$.}
In total, $M=6$ particle sizes are available with particle radius sizes in the range from \SIrange{10}{110}{\nano\meter}.
The distance between transmitter and receiver is $d=\SI{10}{\micro\meter}$ and the spherical receiver radius is $a=\SI{1}{\micro\meter}$.
The symbol duration is chosen as $T=\SI{120}{\second}$ and 600 samples per symbol are considered for discretization, i.e., the discrete time step is $\Delta t=\SI{0.2}{\second}$.
The detection threshold is $\detectionthreshold=15$ and the ISI threshold is $\interferencethreshold=8$.

\begin{figure}
\centering
\includegraphics[scale=\figscal]{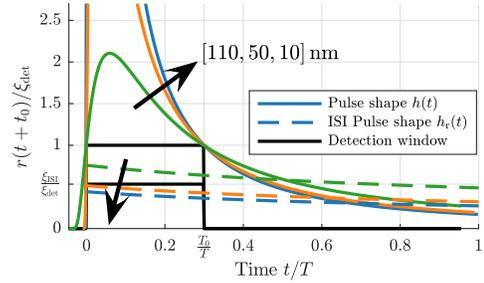}
\vspace*{-3mm}
\caption{%
	Single particle size optimization.
	Worst-case received signals for detecting $a_0=1$ and $a_0=0$ after minimization without ISI constraint.
	Three particle sizes are considered and a detection duration of $T_0=\num{0.25}\cdot T=\SI{30}{\second}$ is chosen.
	For better comparison, all signals are shifted by $t_0$ to align with the detection threshold at time $t=0$.
	The \maxsignal{} for the considered particle sizes $R=\SIlist{10; 50; 110}{\nano\meter}$ is $N=\numlist{3.2e6; 3.4e5; 1e5}$, respectively.
	\vspace*{-5mm}
}
\copyablelabel{fig:illustration}
\end{figure}
\begin{figure*}
\centering
\includegraphics[scale=\figscal]{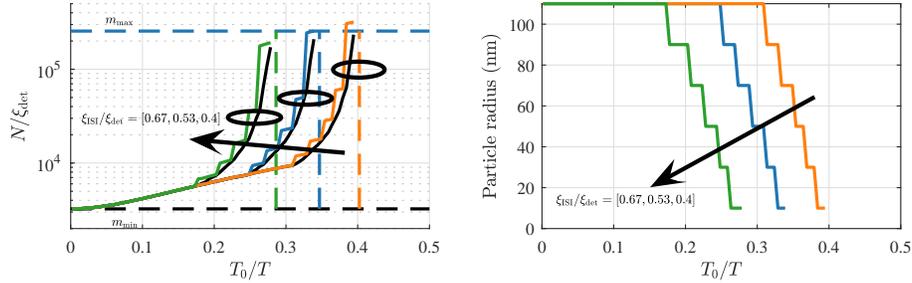}
\vspace*{-3mm}
\caption{%
	Optimization results.
	Peak detection value versus detection duration (left-hand side) and optimal single particle size versus detection duration (right-hand side).
	Optimization results using all available particle sizes (black lines) and using the largest feasible particle size for each detection duration (blue, orange, and green lines) are shown.
	Results are evaluated for three different ISI-threshold-to-detection-threshold ratios.
	Dashed lines show the theoretical minimum and maximum \maxsignal{} as well as the maximum feasible detection duration in the small particle limit.
	\vspace*{-5mm}
}
\copyablelabel{fig:results}
\end{figure*}
%

In Fig.~\ref{fig:illustration}, we illustrate the receive pulse shapes in \eqref{eq:pulse shape} obtained by the optimization in \eqref{eq:optimization} for individual particle sizes.
The y-axis is cut off for clear presentation of the results.
For the time signals, we show the detection lower bound $\detectionthreshold$, as well as the ISI upper bound $\interferencethreshold=\num{0.53}\detectionthreshold$.
We consider three cases, namely using only the smallest available particle size $\SI{10}{\nano\meter}$, only the largest available particle size $\SI{110}{\nano\meter}$, and an intermediate particle size $\SI{50}{\nano\meter}$.
For the largest considered particle size, we adopted $\interferencethreshold=\detectionthreshold=15$ for illustration since for $\interferencethreshold=8$ no solution exists.
Results are obtained by solving the optimization problem in \eqref{eq:optimization}.
We can observe that all received signals closely satisfy the detection constraint, i.e., the time span where the received signal is above the detection threshold matches the prescribed detection duration $T_0=\num{0.3}T$.
This is expected since we are minimizing the \maxsignal{}.
Moreover, for the intermediate particle size, the worst-case ISI signal also intersects with the ISI threshold, i.e., the inequality constraints are met with equality.
Using only the smallest particle size is feasible but requires a lot of resources and produces a large overshoot of the signal with respect to the detection threshold.
On the other hand, using only the largest particle size minimizes the particle usage while the received signal lies above the detection threshold but is actually infeasible since the worst-case ISI signal lies above the ISI threshold.
An intermediate particle size can ensure that the received signal lies above the detection threshold and below the ISI threshold while the \maxsignal{} is minimized at the transmitter.
\changed{In this way, MC using an appropriately chosen particle size can be expected to not exceed a given symbol error rate for a wide range of timing offsets.}

In Fig.~\ref{fig:results}, we evaluate the performance of the proposed optimization scheme.
To this end, we show the optimized \maxsignal{} as a function of the detection duration (indicating robustness to detection time offsets) on the left-hand side.
On the right-hand side, we show the particle selection for a benchmark where just a single particle size is used which is chosen among all available particle sizes such as to minimize the \maxsignal{}.
For the optimization, we consider ISI thresholds $\interferencethreshold=8$ (baseline), $\interferencethreshold=6$, and $\interferencethreshold=10$.
Moreover, we show the lower bound on the \maxsignal{} in \eqref{eq:minimumresource} as well as the maximum feasible detection duration for very small particles in \eqref{eq:smallparticlebound}.
For the baseline, we also show the upper bound on the \maxsignal{} in \eqref{eq:smallparticlescaling}.
\changed{As can be observed, all bounds are asymptotically tight.}
For small detection durations, the \maxsignal{} is identical for different ISI thresholds since here the largest particle size is chosen, see also the right-hand side, i.e., the ISI constraint does not yet play a role and using only the largest particle is optimal.
For larger detection durations, the \maxsignal{} is lower for the combination of all particle sizes compared to the benchmark of selecting just a single particle size.
The maximum supported detection duration is virtually identical for using all particle sizes compared to the benchmark which for large detection durations only selects the smallest available particle size.
For system design, we can conclude that for small detection durations large particles are suitable while for large detection windows relatively small particles are needed due to the ISI constraint.
Overall, a combination of all available particle sizes outperforms using just a single particle size.
However, a suitable selection \changed{among} finitely many single particle sizes achieves reasonably good results.

\section{Conclusions}
\label{sec:conclusion}
In this paper, we have proposed a \changed{method} for pulse shaping based on using the different propagation characteristics of differently sized particles.
We have optimized the particle size mixture at the transmitter for guaranteeing both a minimum signal for detecting the presence of a signal as well as a maximum signal value for detecting the absence of a signal in a prescribed detection time window \changed{to achieve} robustness against detection time offsets.
For the considered free space scenario, our results indicate optimality of the largest particle size in terms of the \maxsignal{} but with a limited feasible detection window due to increased ISI, and maximum feasibility for the smallest particle size.

\changed{The proposed general optimization framework can be easily extended to other system models, including experimental systems, as long as the individual CIRs for all available particle types are known and a linear superposition of these CIRs is observed at the receiver.}

\ifCLASSOPTIONcaptionsoff
  \newpage
\fi



\bibliographystyle{IEEEtran}
\bibliography{main.bbl}

\begin{thebibliography}{1}
\providecommand{\url}[1]{#1}
\csname url@samestyle\endcsname
\providecommand{\newblock}{\relax}
\providecommand{\bibinfo}[2]{#2}
\providecommand{\BIBentrySTDinterwordspacing}{\spaceskip=0pt\relax}
\providecommand{\BIBentryALTinterwordstretchfactor}{4}
\providecommand{\BIBentryALTinterwordspacing}{\spaceskip=\fontdimen2\font plus
\BIBentryALTinterwordstretchfactor\fontdimen3\font minus
  \fontdimen4\font\relax}
\providecommand{\BIBforeignlanguage}[2]{{%
\expandafter\ifx\csname l@#1\endcsname\relax
\typeout{** WARNING: IEEEtran.bst: No hyphenation pattern has been}%
\typeout{** loaded for the language `#1'. Using the pattern for}%
\typeout{** the default language instead.}%
\else
\language=\csname l@#1\endcsname
\fi
#2}}
\providecommand{\BIBdecl}{\relax}
\BIBdecl

\bibitem{jamali_channel_2019}
V.~Jamali, A.~Ahmadzadeh \emph{et~al.}, ``Channel modeling for diffusive
  molecular communication -- {{A}} tutorial review,'' \emph{Proc. IEEE}, vol.
  107, no.~7, pp. 1256--1301, Jul. 2019.

\bibitem{kuran_survey_2021}
M.~{\c S}. Kuran, H.~B. Yilmaz \emph{et~al.}, ``A survey on modulation
  techniques in molecular communication via diffusion,'' \emph{IEEE Commun.
  Surv. Tutor.}, vol.~23, no.~1, pp. 7--28, 1st quart. 2021.

\bibitem{wyatt_pheromones_2014}
T.~D. Wyatt, \emph{Pheromones and {{Animal Behavior}}: {{Chemical Signals}} and
  {{Signatures}}}.\hskip 1em plus 0.5em minus 0.4em\relax {Cambridge University
  Press}, Jan. 2014.

\bibitem{zaloga_development_2014}
J.~Zaloga, C.~Janko \emph{et~al.}, ``Development of a lauric acid/albumin
  hybrid iron oxide nanoparticle system with improved biocompatibility,''
  \emph{Int. J. Nanomed.}, vol.~9, pp. 4847--4866, Oct. 2014.

\bibitem{jamali_symbol_2017}
V.~Jamali, A.~Ahmadzadeh, and R.~Schober, ``Symbol synchronization for
  diffusion-based molecular communications,'' \emph{IEEE Trans.
  NanoBioscience}, vol.~16, no.~8, pp. 873--887, Dec. 2017.

\bibitem{cao_diffusive_2019}
T.~N. Cao, A.~Ahmadzadeh \emph{et~al.}, ``Diffusive mobile {{MC}} with
  absorbing receivers: {{Stochastic}} analysis and applications,'' \emph{IEEE
  Trans. Mol. Biol. Multi-Scale Commun.}, vol.~5, no.~2, pp. 84--99, Nov. 2019.

\bibitem{farsad_molecular_2016}
N.~Farsad and A.~Goldsmith, ``A molecular communication system using acids,
  bases and hydrogen ions,'' in \emph{Proc. {{IEEE SPAWC}}}, Jul. 2016, pp.
  1--6.

\bibitem{tepekule_novel_2015}
B.~Tepekule, A.~E. Pusane \emph{et~al.}, ``A novel pre-equalization method for
  molecular communication via diffusion in nanonetworks,'' \emph{IEEE Commun.
  Lett.}, vol.~19, no.~8, pp. 1311--1314, Aug. 2015.

\bibitem{angjo_asymmetrical_2022}
J.~Angjo, A.~E. Pusane \emph{et~al.}, ``Asymmetrical relaying in molecular
  communications,'' \emph{IEEE Trans. NanoBioscience}, vol.~21, no.~4, pp.
  570--574, Oct. 2022.

\end{thebibliography}
%

%

%





\end{document}